\documentclass[aps,twocolumn,groupeaddress,prb]{revtex4}

\renewcommand{\vec}[1]{\mbox{\boldmath$#1$}}

\newcommand{\me}{\mathrm{e}}

\newcommand{\dif}{\mathrm{d}}

\newcommand{\pmf}{\emph{pmf\/}}

\usepackage{graphicx}

\usepackage{amsmath}

\usepackage{amsfonts}

\usepackage{graphicx}
\begin{document}
\title{Role of attractive methane-water interactions in the potential of
mean force between methane molecules in  water}
\author{D. Asthagiri}\thanks{Corresponding author: Fax: +1-410-516-5510; Email: dilipa@jhu.edu}
\author{Safir Merchant}
\affiliation{Chemical and Biomolecular Engineering, Johns Hopkins
University, Baltimore, MD 21218}
\author{Lawrence R. Pratt}
\affiliation{Department of Chemical and Biomolecular Engineering, Tulane
University,
New Orleans, LA 70118}
\date{\today}
\begin{abstract}
On the basis of a  gaussian quasi-chemical model of hydration, a model
of {\em non\/} van der Waals character, we explore the role of
attractive methane-water interactions in the hydration of methane and in
the potential of mean force between two methane molecules in water. We
find that the hydration of methane is dominated by packing and a
mean-field energetic contribution. Contributions beyond the mean-field
term are unimportant in the hydration phenomena for a hydrophobic solute
such as methane. Attractive  solute-water interactions make a net
\emph{repulsive} contribution to these pair potentials of mean force.
With no conditioning, the observed distributions of binding energies are
\emph{super}-gaussian and can be effectively modeled by a Gumbel
(extreme value) distribution. This further supports the view that the
characteristic form of the unconditioned distribution in the
high-$\varepsilon$ tail is due to energetic interactions with  a small
number of molecules. Generalized extreme value distributions also
effectively model the results with minimal conditioning, but in those
cases the distributions are sufficiently narrow that details of their shape
aren't significant.
\end{abstract}

\maketitle

\section{Introduction}

It is generally accepted that hydrophobic effects are an essential
contribution to the stability of the functional structures of soluble
proteins.\cite{Tanford:97} After Frank and Evans\cite{FRANKHS:FREVAE}
and then Kauzmann,\cite{KAUZMANNW:SOMFIT} hydrophobic effects have been
identified by distinctive unfavorable entropies of hydration and large
hydration heat capacities.  That has made development of a fully
defensible molecular theory of hydrophobic effects an outstanding
challenge for statistical thermodynamics of solutions.

Over the past decade, however, such a defensible molecular theory has
begun to take shape.\cite{PrattLR:Molthe,AshbaughHS:Colspt} Those
developments\cite{PohorilleA:CAVIML,PrattLR:THEOHT,HummerG:Anitm,PrattLR:HYDEFC} 
started principally from the scaled-particle models,\cite{Stillinger:73}
but also defined a connection to the Pratt-Chandler
theory.\cite{PRATTLR:Thehe,CHANDLERD:Gaufmf,PERCUSJK:Stampc}  In view of
those developments, consolidation of previous work is to be
expected.\cite{PrattLR:Molthe}  This paper addresses the contributions
of attractive interactions, \emph{e.g.,} dispersion interactions, on the
potentials of the average forces between small hydrophobic solutes in
water. That is a topic of specific interest that has been left
unresolved.

Much of that specific interest is  associated with the
Pratt-Chandler theory which was the first serious theoretical prediction
of the potential of mean force, \pmf, between simple hydrophobic solutes
in water.\cite{PrattLR:HYDEFC} A paper\cite{PrattLR:Effsaf} which
followed the initial Pratt-Chandler predictions treated the role of
attractive solute-solvent interactions, and addressed an apparent
disagreement  with experiment that had surfaced.\cite{RosskyPJ:BENIIA}
By now, a significant body of simulation data has accumulated on those 
\pmf s. In the following, we briefly note those results salient to our
goals.  But even more, a recent theoretical
insight\cite{AsthagiriD.:NonWth} seems to provide just the analysis tool
needed to organize and clarify the accumulated simulation data, and to
clarify the theoretical understanding of these \pmf s.

The apparent disagreement  involved the temperature dependence of the
osmotic second virial coefficient
\begin{eqnarray}
B_2 \equiv - \frac{1}{2} \int \left(\me^{-w_{\mathrm{AA}}\left(r\right)/kT} - 1\right) 4\pi r^2 \dif r~,
\label{eq:B2}
\end{eqnarray}
where $w_{\mathrm{AA}}\left(r\right)$ is the \pmf \ between two
hydrophobic solutes A, assumed to be spherical in the original
Pratt-Chandler theory which treated hard-sphere
solutes.  The \emph{sign} of $B_2$ predicted for hard-sphere solutes by
the Pratt-Chandler theory depended on the size of the solute, the
predictions for larger solutes was attractive $B_2<$0. But the magnitude
of $B_2$ decreased with increasing temperature, so the effective attractive
interactions were predicted to become less attractive as the temperature
was raised.  The general expectation for hydrophobic interactions always
had been that $B_2$ should be negative, indicating a net attraction, and
increasing in strength with increasing temperature.   Subsequent
experiments for the non-spherical A = C$_6$H$_6$
(benzene)\cite{TuckerEE:PROHI-,TuckerEE:VAPSOH} and for A = C$_6$F$_6$
(perfluoro-benzene)\cite{BernalP:VAPSOH} were consistent with the
general expectation for hydrophobic interactions: this is a favorable 
\emph{and endothermic} association, \emph{i.e.,} this is an entropy
driven association. Though later results have revised that
endothermicity for benzene association to smaller
values,\cite{HallenD:ENTODO} simulation results on model spherical
solutes in water agree with the prior expectation for hydrophobic
interactions.\cite{LudemannS:Theitp,LudemannS:Thetha}

We note also that study of the benzene-benzene  \pmf \ by direct
numerical simulation has produced results in rough agreement with the
experiments.%
\cite{JORGENSENWL:AROAI-,LINSEP:StaTbd,LINSEP:ORIBBP,ChipotC:BendAg}
These are sufficiently challenging calculations, however, that those
simulation results may not be definitive for $B_2$. A
significant result of those calculations is that the T-shaped
benzene-benzene contacts are more favorable than stacked
configurations.\cite{LINSEP:StaTbd,LINSEP:ORIBBP,ChipotC:BendAg} This
seems to be qualitatively different from the case studied with 
purely repulsive  model plates stacked.\cite{WALLQVISTA:COMOHH} This
detail suggests a general conclusion that attractive solute-solvent
interactions can play a qualitative role in establishing the most
probable contacts.  That realistically modelled benzene dimers and
toluene dimers may  differ significantly in this
regard\cite{ChipotC:BendAg} indicates yet again that realistic details
of the cases considered may decisively affect these contacts .

For spherical hydrophobic solutes, the case of Kr(aq) has received special
attention.  Whether $B_2 >0$ follows from the experimental solubilities 
of Kr(aq) is undecided,\cite{KennanRP:Predsn} but relying on simulation
results again, we do know that  it is a realistic
possibility.\cite{WATANABEK:MOLSOT}  Those simulation results,
furthermore, do suggest that the the sign of $B_2$ can change with the
strength of solute-solvent attractive interactions. With the potential of mean force for a C60 pair,\cite{Li:2005p289} in a
different size regime,  attractive interactions make a dominating
contribution that is \emph{repulsive} in character.

The work of Pangali \emph{et al.},~\cite{PANGALIC:AMCs} 
indicated that the original Pratt-Chandler theory was valid for
$w_{\mathrm{AA}}\left(r\right)$ in the case where the predicted
contribution of attractive interactions was small.\cite{PrattLR:Effsaf}
Furthermore, the theory of Ref.~14
was  observed by Smith \& Haymet~\cite{SMITHDE:Freeea} on the basis of molecular simulation to be
strikingly accurate for $w_{\mathrm{AA}}\left(r\right)$  point-wise in
the Lennard-Jones model case there considered for which the predicted
contribution of attractive interactions was large. For that case
$B_2>0$.\cite{SMITHDE:Freeea}  Nevertheless, the underlying
Pratt-Chandler theory was the source of an inaccurate association of
that Lennard-Jones model with CH$_4$(aq). As an example of the type of
inaccuracy that should be expected, the original Pratt-Chandler theory
is approximate even at small solute sizes for the case of a hard-sphere
solvent.\cite{ChenYG:Diftps}  Thus the accuracy of the \pmf \ result for
that case\cite{SMITHDE:Freeea} was not developed further.

The observations above suggest the following general hypotheses:  (a)
the balance of attractive interactions can play a significant role in hydrophobic
interactions; (b) with attractive interactions included the original
Pratt-Chandler theory can be accurate for the hydrophobic interactions
expressed by $w_{\mathrm{AA}}\left( r \right)$.  The original
Pratt-Chandler theory for hydrophobic \emph{hydration} problems was less
accurate.  But the modern revisions of that theory address those
hydration problems first,\cite{PrattLR:Molthe} and largely on the basis
of scaled-particle approaches\cite{Stillinger:73} which are by now
well-developed.\cite{AshbaughHS:Colspt,AshbaughHS:Connaa} (c) $B_2>0$ is
a realizable possibility, but probably not for the cases of general
experimental interest.

In response to the apparent experimental disagreement noted above, a
variety of complications were considered:  (a) context hydrophobicity
which, for example, might be expressed in experiments on  solutes such
as methanol in aqueous solution.\cite{PrattLR:HYDIAO} (b) non-spherical
shapes of molecules of experimental interest;\cite{PrattLR:Hydsns} and
(c) the influence of attractive solute-solvent attractive
interactions.\cite{PrattLR:Effsaf} These are all valid general concerns.
  But the \emph{post hoc} conclusion has been that the underlying
Pratt-Chandler theory is sufficiently unconvincing that extensions and
elaborations have not been compelling.

Though the intuitive proposal of Ref.~14
for 
inclusion of attractive interactions worked-out
accurately in some cases tested,\cite{SMITHDE:Freeea,PANGALIC:AMCs}
there is a logical complication.  It is common to assume that attractive
interactions may be included perturbatively \emph{after} packing
problems dominated by repulsive solute-solvent interactions are solved. 
Then it is expected that attractive interactions may change
thermodynamic properties significantly while making small
alterations to structural characteristics. Being an integrated quantity,
$B_2$ is a thermodynamic property and should be expected to be sensitive
to inclusion of attractive interactions, from this perspective.  But the
assumed insensitivity of structural characteristics has not been so
carefully examined; a physical comment about the complicated role of
hard-core models of hydrophobic solutes was attempted in
Ref.~5,
and  further observations on the role of attractive interactions in
hydrophobic solubility models were made by Paschek.\cite{PaschekD:Temdhh} 

A first perspective on the complications that can arise in starting from
hard-core model solutes is that those complications are precursors of
the drying that is expected on large length scales.\cite{Stillinger:73}
Drying has been discussed broadly in recent
times.\cite{LumK:Hydsal,WeeksJD:Conlsi}  Similarly, the startling
success of the information theory approach for hard-sphere solutes of
intermediate size\cite{HummerG:Anitm,GomezMA:Molrdm} derives from a
balance of errors, approximate treatment of packing problems balanced by
neglect of incipient drying.\cite{PrattLR:Molthe} An additional
perspective is that the difficulties of the original Pratt-Chandler
theory are associated with detailed problems of Percus-Yevick-analogue
theories as was noted above.\cite{ChenYG:Diftps}    Multiple
complications of different types makes correcting a reference
(hard-core) theory tricky. Nevertheless, the status of the
Pratt-Chandler theory for hard-sphere solutes in water has
changed\cite{PrattLR:Molthe}, and the amended theory is now a compelling
approximate theory --- a quasi-chemical theory\cite{PrattLR:Molthe}  ---
relying  on and acknowledging copious empirical input.%
\cite{HummerG:Anitm,HummerG:Hydems,GomezMA:Molrdm,PrattLR:Thehea,HummerG:Newphe}

With this context, the work of Ref.~16
provides a
key to theoretical assessment of the role of attractive interactions for
these problems.  That theory utilizes a result for the hard-core problem
to address cases with attractive interactions, but the logical status of
the hard-core result is \emph{not} that of a reference system.   There
is no uncontrolled assumption of linear perturbative treatment of
attractive interactions.  Indeed, below we calculate non-linear
contributions of two distinct types beyond non-hard-core interactions
and an intuitively recognizable mean-field contribution, as is discussed
below.  Characterization and testing of this alternative theory for such
a \pmf is thus the target of this paper.

\section{Theory}

We recapitulate the gaussian model of hydration and in the process also
set the notation. Then we consider the extension of the theory to
describe the potential of mean force between two methane molecules in
water. 

On the basis of the inverse form of the potential distribution theorem,
the excess chemical potential of methane(aq), $\mu^{\rm ex}_{\mathrm{M}}$, is
given by
\begin{eqnarray}
\me^{\beta \mu^{\rm ex}_{\mathrm{M}}} = \langle \me^{\beta \varepsilon }\rangle = \int P_{\mathrm{M}}(\varepsilon) \me^{\beta \varepsilon }\; \dif\varepsilon ~   ,
\label{eq:invPDT}
\end{eqnarray}
$\beta = 1/k_{\rm B} T$, where $T$ is the temperature. $\langle \ldots
\rangle$ specifies averaging on the basis of the probability density
function $P_{\mathrm{M}}(\varepsilon) = \langle \delta(\varepsilon - \Delta
U_{\mathrm {M}})\rangle $, which defines the distribution of binding energies of the
solute with the solvent. For a particular configuration, the binding
energy $\Delta U_{\mathrm {M}} = U_{N+1} - U_{N} - U_1$ is the difference in the
potential energy of the whole system ($U_{N+1}$) and the sum of the
(decoupled) potential energies of the solution ($U_{N}$) and the solute
($U_1$). Eq.~\eqref{eq:invPDT}  is independent of the simulation ensemble in the
macroscopic limit  and an excellent approximation for system sizes considered
in this work \cite{lrp:cpms}.

Notice that the exponential weighting exhibited by Eq.~\eqref{eq:invPDT}
emphasizes high binding-energy contributions to
$P_{\mathrm{M}}(\varepsilon)$. These high energy contributions reflect
collisions at short-range  between the solute and the solvent particles. These
high-$\varepsilon$ (low-probability) features also render the direct
application of Eq.~\eqref{eq:invPDT} problematic. 

Instead we \emph{regularize} of the statistical problem as follows:
Consider a hard-core (HC) solute  that excludes water oxygen atoms
from a spherical region of radius $\lambda$ centered on the methane. 
$\mu^{\rm ex}_{\rm HC}$, the excess chemical potential of this hard-core solute, 
is assumed to be known. On the basis of simulation data obtained from a calculation 
on methane-water system with realistic interactions, $\mu^{\rm ex}_{\rm HC}$ is given by
\begin{eqnarray}
\me^{-\beta (\mu^{\rm ex}_{\rm HC} - \mu^{\rm ex}_{\mathrm{M}})} = 
\langle \me^{-\beta(\varepsilon_{\rm HC} - \varepsilon)}\rangle~.
\label{eq:MetoHC}
\end{eqnarray}
$\varepsilon_{\rm HC}$ is the binding energy of the hard-core model solute;
this is  zero when no water molecules overlap the hard-core model solute and
is infinite otherwise.  $\varepsilon$ is the binding energy of
M. The use of Eq.~\eqref{eq:MetoHC} requires eliminating configurations
in which a solvent molecule penetrates the assigned hard-core exclusion
volume in a simulation of M in water. The fraction of configurations
thus selected is $p_{\mathrm{M}}(n_\lambda\! =\! 0) $. Then
\begin{multline}
\beta\mu^{\rm ex}_{\mathrm{M}}  =  \beta \mu^{\rm ex}_{\rm HC}  + \ln p_{\mathrm{M}}(n_\lambda\! =\! 0) 
\\
+  \ln \int P_{\mathrm{M}}(\varepsilon | n_\lambda\! =\! 0) \me^{\beta \varepsilon }\; \dif\varepsilon ~.
\label{eq:regular}
\end{multline}
Here $P_{\mathrm{M}}(\varepsilon | n_\lambda\! =\! 0)$ is the binding energy
distribution conditional on the event that there are no overlaps with
the defined hard-core model solute.  The virtue of the regularization
is that the troublesome high-$\varepsilon$ features are accounted
for by $\mu^{\rm ex}_{\rm HC}$ and $p_{\mathrm{M}}(n_\lambda\! =\! 0)$. 
The process is one of pushing the water molecules away from the solute. Thus
by controlling the spatial distribution of molecules we temper the binding energies.  The binding energies remaining after regularization are less energetic.  They are  composed of
contributions from numerous sources, distant and thus weakly
correlated. It is expected that $P_{\mathrm{M}}(\varepsilon | n_\lambda\! =\!
0)$ will be well described by a normal  distribution. If
$P_{\mathrm{M}}(\varepsilon | n_\lambda\! =\! 0)$  is modelled as a Gaussian
distribution, then
\begin{multline}
\beta\mu^{\rm ex}_{\mathrm{M}}  \approx  \beta \mu^{\rm ex}_{\rm HC}  + \ln p_{\mathrm{M}}(n_\lambda\! =\! 0) 
\\
+ \beta \langle \varepsilon | n_\lambda\! =\! 0 \rangle + \frac{\beta^2}{2} \langle  \delta \varepsilon ^2 | n_\lambda\! =\! 0\rangle ~.
\label{eq:gaussian}
\end{multline}
This approach has been tested on the  case of  hydrophobic
CF$_4$(aq)\cite{AsthagiriD.:NonWth} and on the case of  liquid
water in more than one way. \cite{PaliwalA:Anamp,shah:144508}  The
balance between the packing and chemical contributions to
Eq.~\eqref{eq:gaussian}, $\mu^{\rm ex}_{\rm HC}$ and $k_{\mathrm{B}}
T\ln p_{\mathrm{M}}(n_\lambda\! =\! 0)$, respectively,  and particularly with changes
in the  volume of the defined inner-shell has been considered explicitly.   A detailed point of note is
that if the inner-shell is chosen small enough, then $\ln
p_{\mathrm{M}}(n_\lambda\! =\! 0)\approx$ 0, but the packing contribution $\mu^{\rm
ex}_{\rm HC}$ is non-zero.  Then as the volume of the inner-shell is
reduced, the packing contribution should be taken to be the \emph{largest}
value consistent with $\ln p_{\mathrm{M}}(n_\lambda\! =\! 0)\approx$ 0 to sufficient
accuracy.   Thus, for example, with this prescription the gaussian estimate of the
free energy of K$^+$(aq)\cite{AsthagiriD.:Rolfsm} can be slightly but
distinctly improved by adding the appropriate packing contribution.

With the free energy Eq.~\eqref{eq:gaussian} in hand, the entropy
can be accessed through the thermodynamic identity
\begin{multline}
T\left(\frac{\partial S}{\partial n_{\mathrm{M}}}\right)_{T,p,n_{\mathrm{W}}}
= \left(\frac{\partial \left\langle E\right\rangle}{\partial n_{\mathrm{M}}}\right)_{T,p,n_{\mathrm{W}}}
\\
+ p\left(\frac{\partial \left\langle V\right\rangle}{\partial n_{\mathrm{M}}}\right)_{T,p,n_{\mathrm{W}}}
- \mu_{\mathrm{M}}~.
\end{multline}
The ideal contributions to these thermodynamic quantities aren't
specifically interesting, so we subtract the corresponding relation
that obtains when the interaction potential energies vanish.  Recalling that
\begin{eqnarray}
\left\lbrack\left(\frac{\partial \left\langle V\right\rangle}{\partial n_{\mathrm{M}}}\right)_{T,p,n_{\mathrm{W}}}\right\rbrack_{\mathrm{ideal}} = \frac{k_{\mathrm{B}}T}{p}~,
\label{eq:pmvIdeal}
\end{eqnarray}
yields
\begin{multline}
T\left(\frac{\partial S^{\mathrm{ex}}}{\partial n_{\mathrm{M}}}\right)_{T,p,n_{\mathrm{W}}}
= \left(\frac{\partial \left\langle U\right\rangle}{\partial n_{\mathrm{M}}}\right)_{T,p,n_{\mathrm{W}}}
\\
+ p\left(\frac{\partial \left\langle V\right\rangle}{\partial n_{\mathrm{M}}}\right)_{T,p,n_{\mathrm{W}}}
- k_{\mathrm{B}}T
- \mu^{\mathrm{ex}}_{\mathrm{M}}~.
\label{eq:exS}
\end{multline}
Here $\left\langle U\right\rangle$ is the expected value of the
potential energy of the system.  The contribution of the partial molar
energy involves the mean binding energy --- without the conditioning
discussed above --- but also the alteration of solvent-solvent
interactions due to the presence of the solute.  The significance of
this term can be appreciated by considering the case of hard-sphere
model methane.   Then the theory Eq.~\eqref{eq:gaussian} without the
outer-shell interaction contributions is transparently correct, and the
partial molar energy contribution to Eq.~\eqref{eq:exS} involves only
changes in solvent-solvent interactions.

Similarly, it is helpful to discuss the contribution to
Eq.~\eqref{eq:exS} associated with the partial molar volume. The
magnitudes of solute partial molar volumes for hydrophobic solutes are
similar to, typically somewhat smaller than, molecular van der Waals
volumes. For moderate pressures and in view of the the typical low
solvent compressibility, the contribution from the actual partial molar
volume is negligible.  The net contribution from the \emph{excess}
partial molar volume in Eq.~\eqref{eq:exS} amounts invariably to $-
k_{\mathrm{B}}T$, the specific value of the full partial molar volume in
Eq.~\eqref{eq:exS} being numerically irrelevant.

\section{Methane-methane potential of mean force}

Now let us consider how to use this statistical thermodynamic model to
analyze the methane-methane \pmf.  Consider a first methane molecule
centered at the origin.  In fact, we will use a united-atom
representation of the methane molecule.   Then focus on the
distribution of a second methane molecule in the field of the first.  That
distribution can be analyzed beginning from the principle of constancy
of the chemical potential in non-uniform systems\cite{Beck:2006}
\begin{eqnarray}
\rho_{\mathrm {M}} (\vec{r} ) & = & 
\left(
\frac{
\me^{\beta\mu_{\mathrm{M}}}
}{
\Lambda_{\mathrm{M}}{}^3
}
\right)
\me^{- \beta u_{\mathrm {MM}}(r)}
\left\langle
\me^{-\beta\Delta U_{\mathrm {M}} } \vert 
\vec{r} \right\rangle _0 \nonumber \\
 &= & \rho_{\mathrm {M}} \me^{- \beta u_{\mathrm {MM}}(r)}\left(
 \frac{\left\langle
\me^{-\beta\Delta U_{\mathrm {M}} } \vert 
\vec{r} \right\rangle _0}{\left\langle
\me^{-\beta\Delta U_{\mathrm {M}} } \right\rangle _0} 
 \right)~.
\label{rho-3}
\end{eqnarray}
Here $u_{\mathrm {MM}}(r)$ is the assumed M-M pair interaction potential
energy function, and $\Lambda_{\mathrm{M}}$ is the thermal deBroglie
wavelength for a methane molecule.  The brackets $\left\langle \ldots
\right\rangle _0$ indicate a \emph{test particle} average, \emph{i.e.}
averaging in the absence of coupling between the medium and the second M
molecule.  $\Delta U_{\mathrm {M}}$ is the binding energy of
the second M molecule with the rest of the solvent medium. 
The conditional  test-particle average $\left\langle \ldots
\vert \vec{r}\right\rangle _0$ is for the second M at position
$\vec{r}$.  The unconditional average will be practically equal to the
conditional average for positions $\vec{r}$ well separated from the
first M molecule at the origin.  The replacement
\begin{eqnarray}
\left\langle
\me^{-\beta\Delta U_{\mathrm {M}} } \vert 
\vec{r} \right\rangle _0 \dif^3 r \rightarrow \left\langle
\me^{-\beta\Delta U_{\mathrm {M}} } \vert r \right\rangle _0  4 \pi r^2 \dif r
\end{eqnarray}
for radial characteristics of the density profile provides the desired 
radial distribution function $g_{\mathrm{MM}}\left(r\right)$ according to 
\begin{eqnarray}
g_{\mathrm{MM}}\left(r\right) = \me^{- \beta u_{\mathrm {MM}}(r)}\left(
 \frac{\left\langle
\me^{-\beta\Delta U_{\mathrm {M}} } \vert r\right\rangle _0}{\left\langle
\me^{-\beta\Delta U_{\mathrm {M}} } \right\rangle _0} 
 \right)~.
\label{eq:g0}
\end{eqnarray}
The inverse form\cite{Beck:2006} corresponding to
Eq.~\eqref{eq:g0} is 
\begin{eqnarray}
g_{\mathrm{MM}}\left(r\right) = \me^{- \beta u_{\mathrm {MM}}(r)}\left(
 \frac{\left\langle
\me^{\beta\Delta U_{\mathrm {M}} } \right\rangle}{
\left\langle
\me^{\beta\Delta U_{\mathrm {M}} } \vert r\right\rangle} 
 \right)~.
\label{eq:g1}
\end{eqnarray}
In this case, the second M molecule is physically present at the radius $r$
in carrying-out the average and thus the subscript zero (0) is absent.
The numerator in Eq.~\eqref{eq:g1} is the  average for a solitary M
molecule, or for the second M molecule infinitely removed from the first
one located at the origin.  The desired \pmf is then
\begin{eqnarray}
-kT \ln g_{\mathrm{MM}}\left(r\right) 
=    u_{\mathrm {MM}}(r)  +  \Delta w_{\mathrm {MM}}(r) ,
\label{eq:pmf}
\end{eqnarray}
which defines $\Delta w_{\mathrm {MM}}(r)$, and then further
denotes
\begin{eqnarray}
 \Delta w_{\mathrm {MM}}(r) + \mu^{\mathrm{ex}}_{\mathrm{M}}
 = kT \ln \left\langle
\me^{\beta\Delta U_{\mathrm {M}} } \vert r\right\rangle~.
\label{eq:defpmf}
\end{eqnarray}

Following the steps that lead to Eq.~\eqref{eq:gaussian}, here we condition simulation data for
the case of two actual M  present and obtain
\begin{multline}
\Delta w_{\mathrm {MM}}(r) + \mu^{\rm ex}_{\mathrm{M}}  \approx  
 \mu^{\rm ex}_{\rm HC}(r)  
     + k_{\mathrm{B}}T \ln p_{\mathrm{M}}( n_\lambda\! =\! 0|r)  \\
      +   \langle \varepsilon \vert 
r, n_\lambda\! =\! 0 \rangle + \langle  \delta \varepsilon^2 \vert 
r, n_\lambda\! =\! 0\rangle /2kT~.
\label{eq:theory}
\end{multline}
(The averaging is over a probability density distribution $P_{\mathrm{M}}(\varepsilon|r) = \langle \delta(\varepsilon - \Delta U_{\mathrm {M}}) | r \rangle $.) 
These formulae apply to the second M molecule physically present at
radius $r$ with the first M molecule located at the origin, and
the indicated binding energies $\varepsilon$ are for the second M
molecule. Nevertheless,
\begin{eqnarray}
\mu^{\rm ex}_{\rm HC}(r) \equiv -k_{\mathrm{B}}T\ln p\left(n_\lambda\!=\!0|r\right)
\label{eq:HC}
\end{eqnarray}
interrogates spontaneous molecular scale cavity formation in the
neighborhood of a single M molecule, the first M molecule which is
located at the origin.

Eq.~\eqref{eq:theory} is clearly approximate, but, as in Eq.~\ref{eq:gaussian}, the binding
energy distribution of only one M molecule is regularized. Further it is a physically motivated theory and also 
systematically organized.  Thus, it has
the advantages of physical theories, \emph{e.g.} providing physical
explanations of the observed behaviors.  More specifically it is
directly associated with the physical theory Eq.~\eqref{eq:gaussian}, so
this approach also offers an internal calibration, in contrast to
calculations of mean forces  that determine the spatially varying
\pmf s but not a constant of integration.

\section{Methods}

We utilize computational results from simulations involving zero,
one, and two methane molecules.  A united atom representation of methane
(Lennard-Jones $\varepsilon\! =\! 0.294$~kcal/mole and $\sigma = 3.73$~{\AA})
was adopted.  The SPC/E water model was used in all the simulations 
with the rigid H$_2$O structure of each water molecule controlled by the
SHAKE method.

Classical molecular dynamics simulations were performed with the NAMD
2.6 code with constant NPT procedures. The pressure was held constant at
1 bar using the Langevin piston method with the Langevin piston period
set at 200 fs with a decay constant of 100 fs.  The temperature of
298~K was controlled by Langevin dynamics applied to the oxygen atoms. 
The temperature of the Langevin piston was set equal to the temperature
of the Langevin thermostat. The damping coefficient of the thermostat
was set at 1 ps$^{-1}$.

A simulation of pure water  was conducted for a system of 512 water
molecules. The Lennard-Jones interaction was terminated at 10.5~{\AA} by
smoothly switching to zero starting at 10~{\AA}. Electrostatic
interactions were treated with the particle mesh Ewald method with a
grid spacing of approximately 0.75~{\AA}.   A well-equilibrated box of
water molecule was used and a further equilibration for 250~ps was
conducted. The production run lasted 750~ps and configurations were
saved every 0.2~ps for analysis.  $\beta\mu^{\rm ex}_{\rm HC}$ was
evaluated by trial insertions into liquid water.  Similarly, we
estimated the test particle distribution
$P^{(0)}_{\mathrm{M}}(\varepsilon) = \langle \delta(\varepsilon - \Delta
U)\rangle_0$ on the basis of about 2.7~million trial insertions of the
methane into these water systems.

\begin{figure}[h!]
\includegraphics[width=3.2in]{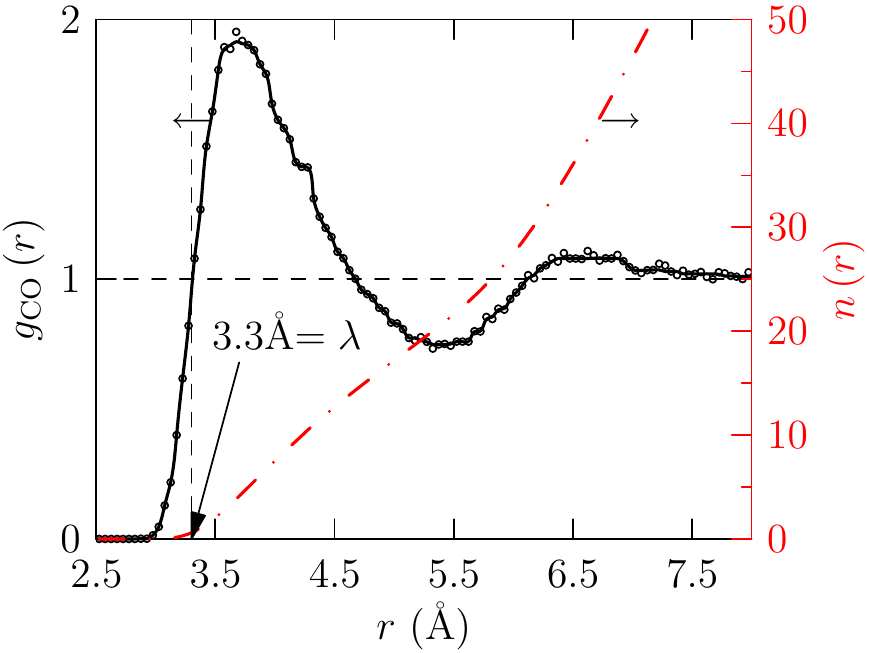}
\caption{The radial distribution function of water oxygen atoms around
methane. 3.3~{\AA}, the smallest distance for which
$g_{\mathrm{CO}}\left(r\right) = 1$, is chosen as the radius $\lambda$
of the inner shell.}
\label{fg:MeO}
\end{figure}

For the methane in water simulation, a methane molecule was placed in
the center of the water box. For convenience in analysis, the methane
was held fixed. The system was extensively equilibrated for over 2~ns.
The production run lasted 4~ns and configurations were saved every
0.5~ps. FIG.~\ref{fg:MeO} shows the CO radial distribution function
observed.

\begin{figure}[h!]
\includegraphics[width=3.2in]{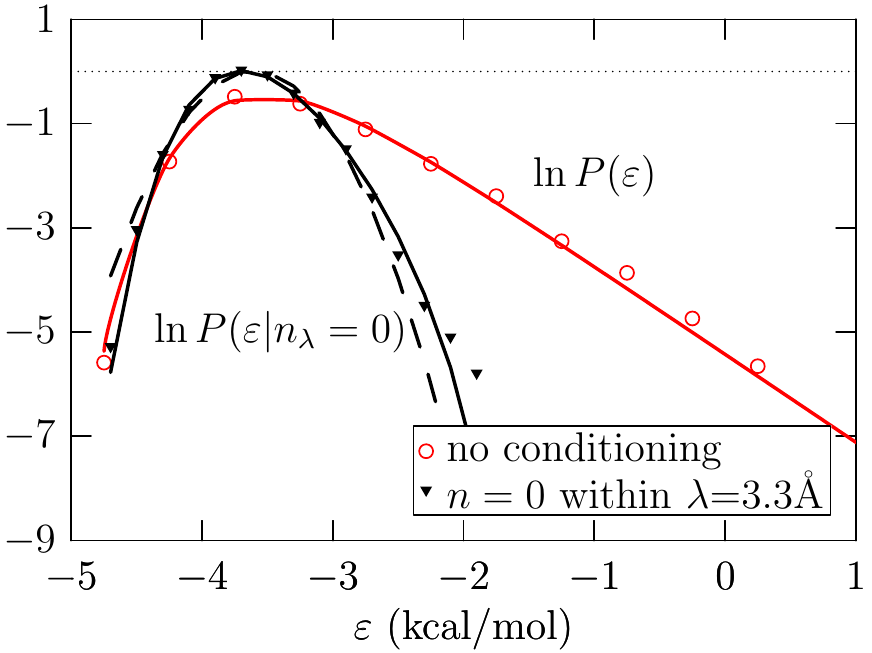}
\caption{Distribution of binding energies of methane to water. The open
circles are the unconditional distribution
$P_{\mathrm{M}}(\varepsilon)$. The filled circles are the conditional
distribution $P_{\mathrm{M}}(\varepsilon | n_\lambda\! =\! 0)$ obtained
from the sample in which no water O-atoms occupy the volume defined by
distances $ \leq 3.3$ {\AA} from the methane center. The dashed curve
is a gaussian model for the conditional distribution. The solid curves
are generalized extreme value\cite{castillo2005} distributions fitted to
the data. For the conditioned distribution this is the Frechet extreme
value distribution $\ln P(\varepsilon)$ = $(n-1)  \ln\left(1 -
(\frac{\varepsilon-a}{n b})\right) - \left(1 - (\frac{\varepsilon-a}{n
b})\right)^n - \ln b$ with $n$ = 6.34, $a$ = -3.76, and $b$ = 0.37.  The
model plotted for the unconditioned distribution is the Gumbel extreme
value distribution discussed in the text.}
\label{fg:MeGaussian}
\end{figure}

$P_{\mathrm{M}}(\varepsilon)$ was assessed and its overlap with
$P^{(0)}_{\mathrm{M}}(\varepsilon)$ was used to estimate $\mu^{\rm
ex}_{\mathrm{M}}$ according to the standard relation $k_\mathrm{B}T\ln
P_{\mathrm{M}}(\varepsilon) /  P^{(0)}_{\mathrm{M}}(\varepsilon)=
-\varepsilon + \mu^{\rm ex}_{\mathrm{M}}$.  A conditioning radius of 
3.3~{\AA} was adopted in considering $P_{\mathrm{M}}(\varepsilon | n_\lambda\!
=\! 0)$. Thus only configurations in which no water molecules come
closer than 3.3~{\AA} to the methane are retained for analysis. The same
conditioning radius is used in all the gaussian models below. 
FIG.~\ref{fg:MeGaussian} shows the unconditional and conditional energy
distributions.

To compute $\mu^{\rm ex}_{\rm HC}(r)$ we adopted the following approach.
For each radial distance from the methane, 572 points were placed to
cover a sphere uniformly. Observation spheres of radius 3.3~{\AA} were
placed at each point and the water occupancy statistic compiled.
Statistics were accumulated from 8000 frames for about 4.6~million
sample points. $p\left(0|r\right)$, the probability that the
observation volume is empty immediately gives $\mu^{\rm ex}_{\rm HC}(r)$
according to Eq.~\eqref{eq:HC}.

The theory above was tested against numerically exact results obtained
by an overlap method as follows. Firstly, we obtained  the distribution
$P_{\mathrm{M}}\left(\varepsilon \vert r \right)$ of binding energies of
one of two  methane molecules that are present in a water system, the
two methane molecules being placed along the box diagonal at various
separations, $r$, ranging from 4~{\AA} to 8~{\AA} in $\Delta r$ =
0.2~{\AA} intervals. The molecules were held fixed and the system
equilibrated for over 1~ns. A production run 750~ps long was conducted
with configurations saved every 0.2~ps. Secondly, we obtained the
distribution $P^{(0)}_{\mathrm{M}}\left(\varepsilon \vert r \right)$ of
a test molecule positioned at $r$ in a corresponding simulation of water
and \emph{one} methane molecule.  These distributions are related by
\begin{eqnarray}
\frac{P_{\mathrm{M}}\left(\varepsilon \vert r \right)
}{
P^{(0)}_{\mathrm{M}}\left(\varepsilon \vert r \right)} 
= \me^{-(\varepsilon - \Delta w_{\mathrm {MM}}(r) - \mu^{\rm ex}_{\mathrm{M}})/kT}~.
\label{eq:overlap}
\end{eqnarray}
Thus, evaluation of these distributions accurately enough to determine this
ratio through an intermediate $\varepsilon$ regime determines $\Delta
w_{\mathrm {MM}}(r) + \mu^{\rm ex}_{\mathrm{M}}$.

\section{Results and Discussion}

FIG.~\ref{fg:MeO} shows the  distribution of water oxygen atoms radially
from the methane carbon. $\lambda = 3.3$~{\AA} is the smallest distance
for which $g_{\mathrm{CO}}\left(\lambda\right) = 1$. Below this radius,
the influence of repulsive interaction dominates the interactions, and
the conditional density of water becomes low. We define the  radius of
the hard sphere to be this value, and the probability $p_M(n_\lambda =
0)$ follows from that.  Here that probability is about $p_M(n_\lambda =
0)\approx 0.6$, so the conditioning eliminates only about half of the total number of configurations.
The \emph{mean} number of water O-atoms with $r<\lambda$ is also about
0.6.  From this we can conclude that $n_\lambda$ = 1, and $n_\lambda$ = 2
are about equally likely with probability of 0.2.

This minimal conditioning, nevertheless, has a dramatic affect on the
distribution of binding energies as FIG.~\ref{fg:MeGaussian} shows. The
extended tail seen there in the unconditional distribution reflects the
high-energy short-range collisions between the methane solute and the
nearest few water molecules, as discussed above.  The
conditioning excludes those high-$\varepsilon$ interactions and the
remaining interactions conform to the anticipated gaussian distribution.
For the conditional distribution $\langle \varepsilon | n_\lambda\!=\!0 \rangle
= -3.6$~kcal/mole, whereas $\langle \varepsilon \rangle =
-3.2$~kcal/mole. Based on the fraction of configurations that contribute
to the conditional distribution, evacuation of the inner-shell requires
the modest energy $-k_{\mathrm{B}} T \ln
p_{\mathrm{M}}(n_\lambda\!=\!0)$ = 0.29~kcal/mole. The latter quantity
is the chemical contribution identified in earlier quasi-chemical
generalizations of the potential distribution
theorem.\cite{AsthagiriD.:NonWth,shah:144508} The fluctuation
contribution $\langle \delta\varepsilon^2 | n_\lambda\!=\!0 \rangle/2kT=
0.13$~kcal/mole. This is about 4\% of the mean binding energy. Thus a
mean-field approximation is adequate to describe the hydration of
methane. By trial insertions, we estimate $\mu^{\mathrm{ex}}_{\rm HC} =
6.08$~kcal/mole, in good agreement with the value of 6.2~kcal/mole
predicted by revised scaled particle theory\cite{AshbaughHS:Colspt}. Thus on the basis of
Eq.~\eqref{eq:gaussian}, $\mu^{\rm ex}_{\mathrm{M}} = 2.32$~kcal/mole,
in excellent agreement with the value of 2.38~kcal/mole obtained on the
basis of the overlap method (see also Ref.~44). 

We used Eq.~\eqref{eq:exS} to obtain a direct evaluation of the entropy
of the M solute.  The remaining difficulty, which is serious, is the
potential for catastrophic loss of precision in performing a subtraction
to evaluate  the partial molar excess internal energy, $\left(\partial
\left\langle U\right\rangle /\partial n_{\mathrm{M}}\right)_{T,p,n_{\mathrm{W}}}$. We
followed\cite{MATUBAYASIN:THEOTH,Lazaridis:2000p304} in considering the
mean binding energies of water molecules within a sphere surrounding the
solute, and seeking saturation and plateau behavior with increasing
radius of the sphere. We note again that the present calculation
permitted the volume of the simulation cell to fluctuate with $p$ (the
pressure) specified, and we also note that the system size for the
present calculation is only 512 molecules. We found, however, the mean
net solvent binding energies to the solvent for water molecules in the
second hydration shell of the solute was roughly zero to within
statistical uncertainties of more than ten kcal/mol.  In the first
hydration shell that mean net solvent binding energy of water to the
solvent was about  one kcal/mol, \emph{positive}, within large
statistical uncertainties but decidedly less than ten kcal/mol. The mean
unconditioned solute-solvent binding energy was -3.2~kcal/mol. The
roughly estimated partial molar energy is thus about -2.0~kcal/mol. This
number is in good agreement with experiment,\cite{ben-naim:2016} and thus the
value for the entropy is satisfactory as well. The first such
\emph{direct} entropy evaluation, for liquid water,\cite{shah:144508}
was strikingly simple, but here the study of the temperature dependence
of  the free energy would be more efficient.  An important physical
point is that this evaluation reinforces the conclusion that
solvent-solvent interactions in the neighborhood of a primitive
hydrophobic solute can be \emph{positive}, a point emphasized recently
by conclusive results for hard-sphere solutes.\cite{AshbaughHS:Colspt}
This point suggests limitations in the classic iceberg or clathrate pictures of
hydrophobic hydration. Of course, such values are expected to be strongly
temperature dependent.

FIG.~\ref{fg:MeGaussian} also shows that extreme value
distributions\cite{castillo2005} are effective in modeling the
distributions $P(\varepsilon)$, particularly the unconditioned distribution. The
high-$\varepsilon$ behavior is faithfully exponential, and the Gumbel
extreme value distribution $\ln P(\varepsilon)$ = $-\left(\varepsilon +
3.52\right)/0.59 + 0.52 - \me^{-\left(\varepsilon + 3.52\right)/0.59}$ is
accurate.  This further supports the view that the characteristic form
of the unconditioned distribution in the high-$\varepsilon$ region is
due to energetic (i.e.\ `extreme') interactions with  a small number of molecules. 

A gamma distribution of interaction energies has
been proposed in a different statistical thermodynamic setting for
closing and integrating a differential equation for an entropic equation
of state.\cite{amadei:1560,Amadei:1996p298} A gamma distribution is
qualitatively satisfactory here but not as accurate as the Gumbel
distribution. As does the present work, those previous studies
emphasized \emph{realizability} in free energy models: free energy
evaluations can benefit from restriction to legitimate probability
distributions.  

A distinction from that preceding
work\cite{amadei:1560,Amadei:1996p298} is that we study here the
distribution of an intensive characteristic, a possibility noted
before.\cite{amadei:1560} Then the central limit theorem is not
available to force the unconditioned distribution toward a normal
(gaussian) form. Alternatively, the present theory naturally presents a
physical control feature, the volume of the defined inner shell, to
drive the observed conditioned distribution toward a normal (gaussian)
form.   With minimal conditioning as here, the conditioned distribution
can be slightly \emph{super}-gaussian.   With aggressive conditioning,
these binding energy distributions can become
\emph{sub}-gaussian.\cite{shah:144508} Then an alternative model,
perhaps a beta distribution, might be more effective.

\begin{figure}[h!]
\includegraphics[width=3.2in]{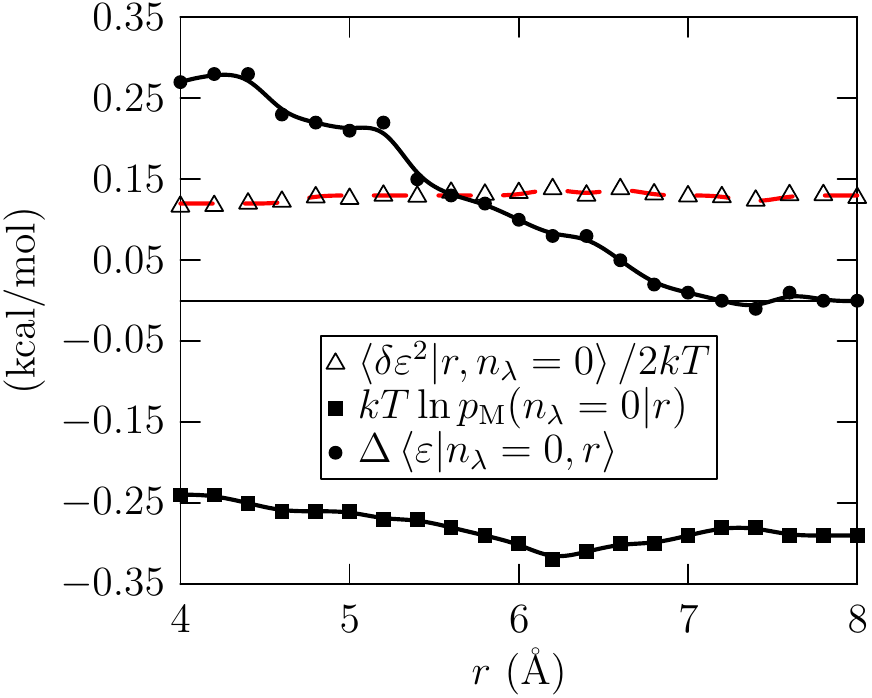}
\caption{Notice the slight variation with radius $r$ of the chemical
contribution $kT\ln p_{\mathrm{M}}( n_\lambda\! =\! 0|r) $ and the fluctuation
contribution $\langle  \delta \varepsilon^2 \vert r, n_\lambda\! =\!
0\rangle/2kT$. The probability that one methane molecule of a pair has
an empty inner-shell ($p_{\mathrm{M}}( n_\lambda\! =\! 0|r)$) is greatest just
outside MM contact (see FIG.~\ref{fg:asym}), and least just outside the
predicted dehydration barrier where intervening water molecules can take
effective advantage of attractive interactions with each methane
molecule of the pair
of methane molecules.}
\label{fg:T0}
\end{figure}

Returning to the results, an important point here is that the chemical
and fluctuation contributions depend weakly on $r$, as FIG.~\ref{fg:T0}
shows.

\begin{figure}[h!]
\includegraphics[width=3.2in]{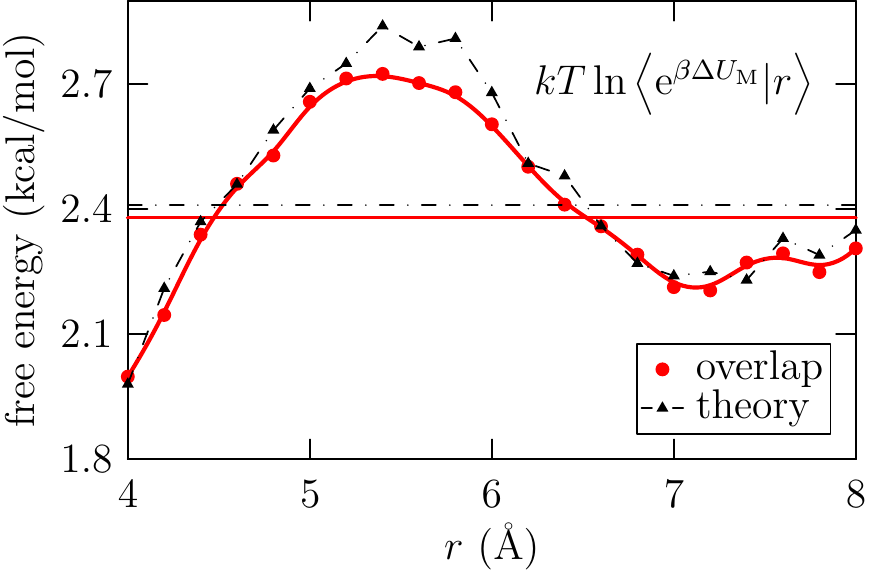}
\caption{The absolute solvent contribution to the potential of mean
force between two methane molecules, see Eq.~\eqref{eq:defpmf}. The bold
line is the reference value obtained from the overlap method of
Eq.~\eqref{eq:overlap}. The triangles are the net profile obtained using
the gaussian-model, Eq.~\eqref{eq:theory}.}
\label{fg:asym} 
\end{figure}

FIG.~\ref{fg:asym} shows $\Delta w_{\mathrm {MM}}(r) +
\mu^{\mathrm{ex}}_{\mathrm{M}}$ (see Eqs.~\eqref{eq:defpmf} and
\eqref{eq:theory}). The agreement between the gaussian quasi-chemical
theory and the numerically exact overlap result is close \emph{except}
in the region of the dehydration barrier. We ascribe the discrepancy of
the theory-overlap data comparison  in that region  to the difficulties
of the present direct estimate of $p\left(n_\lambda\!=\!0|r\right)$ (see
Eq.~\eqref{eq:HC}), particularly in view of the evident noise in the
theoretical result. Together with the results of
FIGS.~\ref{fg:MeGaussian} and \ref{fg:T0}, this is strong confirmation
of the physical correctness of the mean-field description of the effects
of attractive interactions in these \pmf s. FIGS.~\ref{fg:T0} and
\ref{fg:asym} together show that for molecular size solutes, and where
the strength of attractive interactions with water are similar to those
expressed by aliphatic groups, the packing contributions are the biggest
single contribution though mean-field contribution from attractive
interactions are not negligible.  The net effect here of attractive
solute-water interactions is \emph{repulsive}, as was guessed
before,\cite{PrattLR:Effsaf} and we expect that to be the usual case.

Discussion of these points, sharpens our understanding of some earlier
calculations\cite{HummerG:Anitm,Hummer98} that utilized solely packing
contributions for such pair potentials of mean force. In a precisely
defined setting, such models should be augmented to include the effects
of attractive forces. Just as important, information theory models for
composite species that can dissociate have some
peculiarities.\cite{HummerG:Hydems,PrattLR:Molthe} It is likely that
those peculiarities combined with secondary but non-negligible affects
of solute-water attractive interactions are part of the understanding of
the striking initial observations of agreement of the information theory
models with simulation results.\cite{HummerG:Anitm}

\bigskip

\section{Conclusions}

The gaussian quasi-chemical theory is physically valid for these pair
potentials of mean force.  The theory here is sufficiently accurate
that the packing contribution may be inferred from simulation 
data on the system with physical interations. 
In the present approach, the packing
contribution does not have the logical status of a reference system
contribution in a perturbative formulation. In particular, the chemical
contribution thereby defined is non-perturbative. Nevertheless, packing
effects make the largest contribution, and both chemical and fluctuation
contributions are negligible by comparison.    For the methane-methane
pair potential of mean force in water at customary conditions, mean-field
effects associated with solute-water attractive interactions are
secondary but not negligible. Attractive  solute-water  interactions
make a net \emph{repulsive} contribution to these pair potentials of
mean force.  This substantiates a previous physical
guess,\cite{PrattLR:Effsaf} and is consistent with intervening
observations.\cite{SMITHDE:Freeea}  Extreme value distributions accurately
model the distributions observed here, which are \emph{super}-gaussian.

\vfill


\begin{thebibliography}{0}
\expandafter\ifx\csname natexlab\endcsname\relax\def\natexlab#1{#1}\fi
\expandafter\ifx\csname bibnamefont\endcsname\relax
  \def\bibnamefont#1{#1}\fi
\expandafter\ifx\csname bibfnamefont\endcsname\relax
  \def\bibfnamefont#1{#1}\fi
\expandafter\ifx\csname citenamefont\endcsname\relax
  \def\citenamefont#1{#1}\fi
\expandafter\ifx\csname url\endcsname\relax
  \def\url#1{\texttt{#1}}\fi
\expandafter\ifx\csname urlprefix\endcsname\relax\def\urlprefix{URL }\fi
\providecommand{\bibinfo}[2]{#2}
\providecommand{\eprint}[2][]{\url{#2}}

\end{thebibliography}


\begin{thebibliography}{54}
\expandafter\ifx\csname natexlab\endcsname\relax\def\natexlab#1{#1}\fi
\expandafter\ifx\csname bibnamefont\endcsname\relax
  \def\bibnamefont#1{#1}\fi
\expandafter\ifx\csname bibfnamefont\endcsname\relax
  \def\bibfnamefont#1{#1}\fi
\expandafter\ifx\csname citenamefont\endcsname\relax
  \def\citenamefont#1{#1}\fi
\expandafter\ifx\csname url\endcsname\relax
  \def\url#1{\texttt{#1}}\fi
\expandafter\ifx\csname urlprefix\endcsname\relax\def\urlprefix{URL }\fi
\providecommand{\bibinfo}[2]{#2}
\providecommand{\eprint}[2][]{\url{#2}}

\bibitem[{\citenamefont{Tanford}(1997)}]{Tanford:97}
\bibinfo{author}{\bibfnamefont{C.}~\bibnamefont{Tanford}},
  \bibinfo{journal}{Prot. Sci.} \textbf{\bibinfo{volume}{6}},
  \bibinfo{pages}{1358} (\bibinfo{year}{1997}).

\bibitem[{\citenamefont{Frank and Evans}(1945)}]{FRANKHS:FREVAE}
\bibinfo{author}{\bibfnamefont{H.~S.} \bibnamefont{Frank}} \bibnamefont{and}
  \bibinfo{author}{\bibfnamefont{M.~W.} \bibnamefont{Evans}},
  \bibinfo{journal}{J. Chem. Phys.} \textbf{\bibinfo{volume}{13}},
  \bibinfo{pages}{507 } (\bibinfo{year}{1945}).

\bibitem[{\citenamefont{Kauzmann}(1959)}]{KAUZMANNW:SOMFIT}
\bibinfo{author}{\bibfnamefont{W.}~\bibnamefont{Kauzmann}},
  \bibinfo{journal}{Adv. Prot. Chem.} \textbf{\bibinfo{volume}{14}},
  \bibinfo{pages}{1 } (\bibinfo{year}{1959}).

\bibitem[{\citenamefont{Pratt}(2002)}]{PrattLR:Molthe}
\bibinfo{author}{\bibfnamefont{L.~R.} \bibnamefont{Pratt}},
  \bibinfo{journal}{Ann. Rev. Phys. Chem.} \textbf{\bibinfo{volume}{53}},
  \bibinfo{pages}{409 } (\bibinfo{year}{2002}).

\bibitem[{\citenamefont{Ashbaugh and Pratt}(2006)}]{AshbaughHS:Colspt}
\bibinfo{author}{\bibfnamefont{H.~S.} \bibnamefont{Ashbaugh}} \bibnamefont{and}
  \bibinfo{author}{\bibfnamefont{L.~R.} \bibnamefont{Pratt}},
  \bibinfo{journal}{Rev. Mod. Phys.} \textbf{\bibinfo{volume}{78}},
  \bibinfo{pages}{159 } (\bibinfo{year}{2006}).

\bibitem[{\citenamefont{Pohorille and Pratt}(1990)}]{PohorilleA:CAVIML}
\bibinfo{author}{\bibfnamefont{A.}~\bibnamefont{Pohorille}} \bibnamefont{and}
  \bibinfo{author}{\bibfnamefont{L.~R.} \bibnamefont{Pratt}},
  \bibinfo{journal}{J. Am. Chem. Soc.} \textbf{\bibinfo{volume}{112}},
  \bibinfo{pages}{5066 } (\bibinfo{year}{1990}).

\bibitem[{\citenamefont{Pratt and Pohorille}(1992)}]{PrattLR:THEOHT}
\bibinfo{author}{\bibfnamefont{L.~R.} \bibnamefont{Pratt}} \bibnamefont{and}
  \bibinfo{author}{\bibfnamefont{A.}~\bibnamefont{Pohorille}},
  \bibinfo{journal}{Proc. Natl. Acad. Sci USA} \textbf{\bibinfo{volume}{89}},
  \bibinfo{pages}{2995 } (\bibinfo{year}{1992}).

\bibitem[{\citenamefont{Hummer et~al.}(1996)\citenamefont{Hummer, Garde,
  Garcia, Pohorille, and Pratt}}]{HummerG:Anitm}
\bibinfo{author}{\bibfnamefont{G.}~\bibnamefont{Hummer}},
  \bibinfo{author}{\bibfnamefont{S.}~\bibnamefont{Garde}},
  \bibinfo{author}{\bibfnamefont{A.~E.} \bibnamefont{Garcia}},
  \bibinfo{author}{\bibfnamefont{A.}~\bibnamefont{Pohorille}},
  \bibnamefont{and} \bibinfo{author}{\bibfnamefont{L.~R.} \bibnamefont{Pratt}},
  \bibinfo{journal}{Proc. Natl. Acad. Sci. USA} \textbf{\bibinfo{volume}{93}},
  \bibinfo{pages}{8951 } (\bibinfo{year}{1996}).

\bibitem[{\citenamefont{Pratt and Pohorille}(1993)}]{PrattLR:HYDEFC}
\bibinfo{author}{\bibfnamefont{L.~R.} \bibnamefont{Pratt}} \bibnamefont{and}
  \bibinfo{author}{\bibfnamefont{A.}~\bibnamefont{Pohorille}},
  \bibinfo{journal}{Italian Physical Society Conference Proceedings (IPS)}
  \textbf{\bibinfo{volume}{43}}, \bibinfo{pages}{261 } (\bibinfo{year}{1993}).

\bibitem[{\citenamefont{Stillinger}(1973)}]{Stillinger:73}
\bibinfo{author}{\bibfnamefont{F.~H.} \bibnamefont{Stillinger}},
  \bibinfo{journal}{J. Soln. Chem.} \textbf{\bibinfo{volume}{2}},
  \bibinfo{pages}{141} (\bibinfo{year}{1973}).

\bibitem[{\citenamefont{Pratt and Chandler}(1977)}]{PRATTLR:Thehe}
\bibinfo{author}{\bibfnamefont{L.~R.} \bibnamefont{Pratt}} \bibnamefont{and}
  \bibinfo{author}{\bibfnamefont{D.}~\bibnamefont{Chandler}},
  \bibinfo{journal}{J. Chem. Phys.} \textbf{\bibinfo{volume}{67}},
  \bibinfo{pages}{3683 } (\bibinfo{year}{1977}).

\bibitem[{\citenamefont{Chandler}(1993)}]{CHANDLERD:Gaufmf}
\bibinfo{author}{\bibfnamefont{D.}~\bibnamefont{Chandler}},
  \bibinfo{journal}{Phys. Rev. E} \textbf{\bibinfo{volume}{48}},
  \bibinfo{pages}{2898 } (\bibinfo{year}{1993}).

\bibitem[{\citenamefont{Percus}(1993)}]{PERCUSJK:Stampc}
\bibinfo{author}{\bibfnamefont{J.~K.} \bibnamefont{Percus}},
  \bibinfo{journal}{Journal de Physique IV (Colloque)}
  \textbf{\bibinfo{volume}{3}}, \bibinfo{pages}{49 } (\bibinfo{year}{1993}).

\bibitem[{\citenamefont{Pratt and
  Chandler}(1980{\natexlab{a}})}]{PrattLR:Effsaf}
\bibinfo{author}{\bibfnamefont{L.~R.} \bibnamefont{Pratt}} \bibnamefont{and}
  \bibinfo{author}{\bibfnamefont{D.}~\bibnamefont{Chandler}},
  \bibinfo{journal}{J. Chem. Phys.} \textbf{\bibinfo{volume}{73}},
  \bibinfo{pages}{3434 } (\bibinfo{year}{1980}{\natexlab{a}}).

\bibitem[{\citenamefont{Rossky and Friedman}(1980)}]{RosskyPJ:BENIIA}
\bibinfo{author}{\bibfnamefont{P.~J.} \bibnamefont{Rossky}} \bibnamefont{and}
  \bibinfo{author}{\bibfnamefont{H.~L.} \bibnamefont{Friedman}},
  \bibinfo{journal}{J. Phys. Chem.} \textbf{\bibinfo{volume}{84}},
  \bibinfo{pages}{587 } (\bibinfo{year}{1980}).

\bibitem[{\citenamefont{Asthagiri et~al.}(2007)\citenamefont{Asthagiri,
  Ashbaugh, Piryatinski, Paulaitis, and Pratt}}]{AsthagiriD.:NonWth}
\bibinfo{author}{\bibfnamefont{D.}~\bibnamefont{Asthagiri}},
  \bibinfo{author}{\bibfnamefont{H.}~\bibnamefont{Ashbaugh}},
  \bibinfo{author}{\bibfnamefont{A.}~\bibnamefont{Piryatinski}},
  \bibinfo{author}{\bibfnamefont{M.}~\bibnamefont{Paulaitis}},
  \bibnamefont{and} \bibinfo{author}{\bibfnamefont{L.}~\bibnamefont{Pratt}},
  \bibinfo{journal}{J. Am. Chem. Soc.} \textbf{\bibinfo{volume}{129}},
  \bibinfo{pages}{10133 } (\bibinfo{year}{2007}).

\bibitem[{\citenamefont{Tucker and Christian}(1979)}]{TuckerEE:PROHI-}
\bibinfo{author}{\bibfnamefont{E.}~\bibnamefont{Tucker}} \bibnamefont{and}
  \bibinfo{author}{\bibfnamefont{S.}~\bibnamefont{Christian}},
  \bibinfo{journal}{J. Phys. Chem.} \textbf{\bibinfo{volume}{83}},
  \bibinfo{pages}{426 } (\bibinfo{year}{1979}).

\bibitem[{\citenamefont{Tucker et~al.}(1981)\citenamefont{Tucker, Lane, and
  Christian}}]{TuckerEE:VAPSOH}
\bibinfo{author}{\bibfnamefont{E.}~\bibnamefont{Tucker}},
  \bibinfo{author}{\bibfnamefont{E.}~\bibnamefont{Lane}}, \bibnamefont{and}
  \bibinfo{author}{\bibfnamefont{S.}~\bibnamefont{Christian}},
  \bibinfo{journal}{J. Soln. Chem.} \textbf{\bibinfo{volume}{10}},
  \bibinfo{pages}{1 } (\bibinfo{year}{1981}).

\bibitem[{\citenamefont{Bernal et~al.}(1986)\citenamefont{Bernal, Christian,
  and Tucker}}]{BernalP:VAPSOH}
\bibinfo{author}{\bibfnamefont{P.}~\bibnamefont{Bernal}},
  \bibinfo{author}{\bibfnamefont{S.}~\bibnamefont{Christian}},
  \bibnamefont{and} \bibinfo{author}{\bibfnamefont{E.}~\bibnamefont{Tucker}},
  \bibinfo{journal}{J. Soln. Chem.} \textbf{\bibinfo{volume}{15}},
  \bibinfo{pages}{947 } (\bibinfo{year}{1986}).

\bibitem[{\citenamefont{Hallen et~al.}(1988)\citenamefont{Hallen, Wadso,
  Wasserman, Robert, and Gill}}]{HallenD:ENTODO}
\bibinfo{author}{\bibfnamefont{D.}~\bibnamefont{Hallen}},
  \bibinfo{author}{\bibfnamefont{I.}~\bibnamefont{Wadso}},
  \bibinfo{author}{\bibfnamefont{D.}~\bibnamefont{Wasserman}},
  \bibinfo{author}{\bibfnamefont{C.}~\bibnamefont{Robert}}, \bibnamefont{and}
  \bibinfo{author}{\bibfnamefont{S.}~\bibnamefont{Gill}}, \bibinfo{journal}{J.
  Phys. Chem.} \textbf{\bibinfo{volume}{92}}, \bibinfo{pages}{3623 }
  (\bibinfo{year}{1988}).

\bibitem[{\citenamefont{Ludemann et~al.}(1996)\citenamefont{Ludemann,
  Schreiber, Abseher, and Steinhauser}}]{LudemannS:Theitp}
\bibinfo{author}{\bibfnamefont{S.}~\bibnamefont{Ludemann}},
  \bibinfo{author}{\bibfnamefont{H.}~\bibnamefont{Schreiber}},
  \bibinfo{author}{\bibfnamefont{R.}~\bibnamefont{Abseher}}, \bibnamefont{and}
  \bibinfo{author}{\bibfnamefont{O.}~\bibnamefont{Steinhauser}},
  \bibinfo{journal}{J. Chem. Phys.} \textbf{\bibinfo{volume}{104}},
  \bibinfo{pages}{286 } (\bibinfo{year}{1996}).

\bibitem[{\citenamefont{Ludemann et~al.}(1997)\citenamefont{Ludemann, Abseher,
  Schreiber, and Steinhauser}}]{LudemannS:Thetha}
\bibinfo{author}{\bibfnamefont{S.}~\bibnamefont{Ludemann}},
  \bibinfo{author}{\bibfnamefont{R.}~\bibnamefont{Abseher}},
  \bibinfo{author}{\bibfnamefont{H.}~\bibnamefont{Schreiber}},
  \bibnamefont{and}
  \bibinfo{author}{\bibfnamefont{O.}~\bibnamefont{Steinhauser}},
  \bibinfo{journal}{J. Am. Chem. Soc.} \textbf{\bibinfo{volume}{119}},
  \bibinfo{pages}{4206 } (\bibinfo{year}{1997}).

\bibitem[{\citenamefont{Jorgensen and Severance}(1990)}]{JORGENSENWL:AROAI-}
\bibinfo{author}{\bibfnamefont{W.}~\bibnamefont{Jorgensen}} \bibnamefont{and}
  \bibinfo{author}{\bibfnamefont{D.}~\bibnamefont{Severance}},
  \bibinfo{journal}{J. Am. Chem. Soc.} \textbf{\bibinfo{volume}{112}},
  \bibinfo{pages}{4768 } (\bibinfo{year}{1990}).

\bibitem[{\citenamefont{Linse}(1992)}]{LINSEP:StaTbd}
\bibinfo{author}{\bibfnamefont{P.}~\bibnamefont{Linse}}, \bibinfo{journal}{J.
  Am. Chem. Soc.} \textbf{\bibinfo{volume}{114}}, \bibinfo{pages}{4366 }
  (\bibinfo{year}{1992}).

\bibitem[{\citenamefont{Linse}(1993)}]{LINSEP:ORIBBP}
\bibinfo{author}{\bibfnamefont{P.}~\bibnamefont{Linse}}, \bibinfo{journal}{J.
  Am. Chem. Soc.} \textbf{\bibinfo{volume}{115}}, \bibinfo{pages}{8793 }
  (\bibinfo{year}{1993}).

\bibitem[{\citenamefont{Chipot et~al.}(1996)\citenamefont{Chipot, Jaffe,
  Maigret, Pearlman, and Kollman}}]{ChipotC:BendAg}
\bibinfo{author}{\bibfnamefont{C.}~\bibnamefont{Chipot}},
  \bibinfo{author}{\bibfnamefont{R.}~\bibnamefont{Jaffe}},
  \bibinfo{author}{\bibfnamefont{B.}~\bibnamefont{Maigret}},
  \bibinfo{author}{\bibfnamefont{D.~A.} \bibnamefont{Pearlman}},
  \bibnamefont{and} \bibinfo{author}{\bibfnamefont{P.~A.}
  \bibnamefont{Kollman}}, \bibinfo{journal}{J. Am. Chem. Soc.}
  \textbf{\bibinfo{volume}{118}}, \bibinfo{pages}{11217 }
  (\bibinfo{year}{1996}).

\bibitem[{\citenamefont{Wallqvist and Berne}(1995)}]{WALLQVISTA:COMOHH}
\bibinfo{author}{\bibfnamefont{A.}~\bibnamefont{Wallqvist}} \bibnamefont{and}
  \bibinfo{author}{\bibfnamefont{B.~J.} \bibnamefont{Berne}},
  \bibinfo{journal}{J. Phys. Chem.} \textbf{\bibinfo{volume}{99}},
  \bibinfo{pages}{2893 } (\bibinfo{year}{1995}).

\bibitem[{\citenamefont{Kennan and Pollack}(1990)}]{KennanRP:Predsn}
\bibinfo{author}{\bibfnamefont{R.}~\bibnamefont{Kennan}} \bibnamefont{and}
  \bibinfo{author}{\bibfnamefont{G.}~\bibnamefont{Pollack}},
  \bibinfo{journal}{J. Chem. Phys.} \textbf{\bibinfo{volume}{93}},
  \bibinfo{pages}{2724 } (\bibinfo{year}{1990}).

\bibitem[{\citenamefont{Watanabe and Andersen}(1986)}]{WATANABEK:MOLSOT}
\bibinfo{author}{\bibfnamefont{K.}~\bibnamefont{Watanabe}} \bibnamefont{and}
  \bibinfo{author}{\bibfnamefont{H.~C.} \bibnamefont{Andersen}},
  \bibinfo{journal}{J. Phys. Chem.} \textbf{\bibinfo{volume}{90}},
  \bibinfo{pages}{795 } (\bibinfo{year}{1986}).

\bibitem[{\citenamefont{Li et~al.}(2005)\citenamefont{Li, Bedrov, and
  Smith}}]{Li:2005p289}
\bibinfo{author}{\bibfnamefont{L.}~\bibnamefont{Li}},
  \bibinfo{author}{\bibfnamefont{D.}~\bibnamefont{Bedrov}}, \bibnamefont{and}
  \bibinfo{author}{\bibfnamefont{G.~D.} \bibnamefont{Smith}},
  \bibinfo{journal}{J. Chem. Phys.} \textbf{\bibinfo{volume}{123}},
  \bibinfo{pages}{204504} (\bibinfo{year}{2005}).

\bibitem[{\citenamefont{Pangali et~al.}(1979)\citenamefont{Pangali, Rao, and
  Berne}}]{PANGALIC:AMCs}
\bibinfo{author}{\bibfnamefont{C.}~\bibnamefont{Pangali}},
  \bibinfo{author}{\bibfnamefont{M.}~\bibnamefont{Rao}}, \bibnamefont{and}
  \bibinfo{author}{\bibfnamefont{B.~J.} \bibnamefont{Berne}},
  \bibinfo{journal}{J. Chem. Phys.} \textbf{\bibinfo{volume}{71}},
  \bibinfo{pages}{2975 } (\bibinfo{year}{1979}).

\bibitem[{\citenamefont{Smith and Haymet}(1993)}]{SMITHDE:Freeea}
\bibinfo{author}{\bibfnamefont{D.~E.} \bibnamefont{Smith}} \bibnamefont{and}
  \bibinfo{author}{\bibfnamefont{A.~D.~J.} \bibnamefont{Haymet}},
  \bibinfo{journal}{J. Chem. Phys.} \textbf{\bibinfo{volume}{98}},
  \bibinfo{pages}{6445 } (\bibinfo{year}{1993}).

\bibitem[{\citenamefont{Chen and Weeks}(2003)}]{ChenYG:Diftps}
\bibinfo{author}{\bibfnamefont{Y.~G.} \bibnamefont{Chen}} \bibnamefont{and}
  \bibinfo{author}{\bibfnamefont{J.~D.} \bibnamefont{Weeks}},
  \bibinfo{journal}{J. Chem. Phys.} \textbf{\bibinfo{volume}{118}},
  \bibinfo{pages}{7944 } (\bibinfo{year}{2003}).

\bibitem[{\citenamefont{Ashbaugh and Pratt}(2007)}]{AshbaughHS:Connaa}
\bibinfo{author}{\bibfnamefont{H.~S.} \bibnamefont{Ashbaugh}} \bibnamefont{and}
  \bibinfo{author}{\bibfnamefont{L.~R.} \bibnamefont{Pratt}},
  \bibinfo{journal}{J. Phys. Chem. B} \textbf{\bibinfo{volume}{111}},
  \bibinfo{pages}{9330 } (\bibinfo{year}{2007}).

\bibitem[{\citenamefont{Pratt and
  Chandler}(1980{\natexlab{b}})}]{PrattLR:HYDIAO}
\bibinfo{author}{\bibfnamefont{L.~R.} \bibnamefont{Pratt}} \bibnamefont{and}
  \bibinfo{author}{\bibfnamefont{D.}~\bibnamefont{Chandler}},
  \bibinfo{journal}{J. Soln. Chem.} \textbf{\bibinfo{volume}{9}},
  \bibinfo{pages}{1 } (\bibinfo{year}{1980}{\natexlab{b}}).

\bibitem[{\citenamefont{Pratt and
  Chandler}(1980{\natexlab{c}})}]{PrattLR:Hydsns}
\bibinfo{author}{\bibfnamefont{L.~R.} \bibnamefont{Pratt}} \bibnamefont{and}
  \bibinfo{author}{\bibfnamefont{D.}~\bibnamefont{Chandler}},
  \bibinfo{journal}{J. Chem. Phys.} \textbf{\bibinfo{volume}{73}},
  \bibinfo{pages}{3430 } (\bibinfo{year}{1980}{\natexlab{c}}).

\bibitem[{\citenamefont{Paschek}(2004)}]{PaschekD:Temdhh}
\bibinfo{author}{\bibfnamefont{D.}~\bibnamefont{Paschek}}, \bibinfo{journal}{J.
  Chem. Phys.} \textbf{\bibinfo{volume}{120}}, \bibinfo{pages}{6674 }
  (\bibinfo{year}{2004}).

\bibitem[{\citenamefont{Lum et~al.}(1999)\citenamefont{Lum, Chandler, and
  Weeks}}]{LumK:Hydsal}
\bibinfo{author}{\bibfnamefont{K.}~\bibnamefont{Lum}},
  \bibinfo{author}{\bibfnamefont{D.}~\bibnamefont{Chandler}}, \bibnamefont{and}
  \bibinfo{author}{\bibfnamefont{J.~D.} \bibnamefont{Weeks}},
  \bibinfo{journal}{J. Phys. Chem. B} \textbf{\bibinfo{volume}{103}},
  \bibinfo{pages}{4570 } (\bibinfo{year}{1999}).

\bibitem[{\citenamefont{Weeks}(2002)}]{WeeksJD:Conlsi}
\bibinfo{author}{\bibfnamefont{J.~D.} \bibnamefont{Weeks}},
  \bibinfo{journal}{Ann. Rev. Phys. Chem.} pp. \bibinfo{pages}{533 -- 562}
  (\bibinfo{year}{2002}).

\bibitem[{\citenamefont{Gomez et~al.}(1999)\citenamefont{Gomez, Pratt, Hummer,
  and Garde}}]{GomezMA:Molrdm}
\bibinfo{author}{\bibfnamefont{M.~A.} \bibnamefont{Gomez}},
  \bibinfo{author}{\bibfnamefont{L.~R.} \bibnamefont{Pratt}},
  \bibinfo{author}{\bibfnamefont{G.}~\bibnamefont{Hummer}}, \bibnamefont{and}
  \bibinfo{author}{\bibfnamefont{S.}~\bibnamefont{Garde}}, \bibinfo{journal}{J.
  Phys. Chem. B} \textbf{\bibinfo{volume}{103}}, \bibinfo{pages}{3520 }
  (\bibinfo{year}{1999}).

\bibitem[{\citenamefont{Hummer et~al.}(1998{\natexlab{a}})\citenamefont{Hummer,
  Garde, Garc\'{i}a, Paulaitis, and Pratt}}]{HummerG:Hydems}
\bibinfo{author}{\bibfnamefont{G.}~\bibnamefont{Hummer}},
  \bibinfo{author}{\bibfnamefont{S.}~\bibnamefont{Garde}},
  \bibinfo{author}{\bibfnamefont{A.~E.} \bibnamefont{Garc\'{i}a}},
  \bibinfo{author}{\bibfnamefont{M.~E.} \bibnamefont{Paulaitis}},
  \bibnamefont{and} \bibinfo{author}{\bibfnamefont{L.~R.} \bibnamefont{Pratt}},
  \bibinfo{journal}{J. Phys. Chem. B} \textbf{\bibinfo{volume}{102}},
  \bibinfo{pages}{10469 } (\bibinfo{year}{1998}{\natexlab{a}}).

\bibitem[{\citenamefont{Pratt et~al.}(1999)\citenamefont{Pratt, Garde, and
  Hummer}}]{PrattLR:Thehea}
\bibinfo{author}{\bibfnamefont{L.~R.} \bibnamefont{Pratt}},
  \bibinfo{author}{\bibfnamefont{S.}~\bibnamefont{Garde}}, \bibnamefont{and}
  \bibinfo{author}{\bibfnamefont{G.}~\bibnamefont{Hummer}},
  \bibinfo{journal}{NATO ADVANCED SCIENCE INSTITUTES SERIES, SERIES C,
  MATHEMATICAL AND PHYSICAL SCIENCES} \textbf{\bibinfo{volume}{529}},
  \bibinfo{pages}{407 } (\bibinfo{year}{1999}).

\bibitem[{\citenamefont{Hummer et~al.}(2000)\citenamefont{Hummer, Garde,
  Garc\'{i}a, and Pratt}}]{HummerG:Newphe}
\bibinfo{author}{\bibfnamefont{G.}~\bibnamefont{Hummer}},
  \bibinfo{author}{\bibfnamefont{S.}~\bibnamefont{Garde}},
  \bibinfo{author}{\bibfnamefont{A.~E.} \bibnamefont{Garc\'{i}a}},
  \bibnamefont{and} \bibinfo{author}{\bibfnamefont{L.~R.} \bibnamefont{Pratt}},
  \bibinfo{journal}{Chem. Phys.} \textbf{\bibinfo{volume}{258}},
  \bibinfo{pages}{349 } (\bibinfo{year}{2000}).


\bibitem[{\citenamefont{Pratt and Asthagiri}(2007)}]{lrp:cpms}
\bibinfo{author}{\bibfnamefont{L.~R.} \bibnamefont{Pratt}} \bibnamefont{and}
  \bibinfo{author}{\bibfnamefont{D.}~\bibnamefont{Asthagiri}}, in
  \emph{\bibinfo{booktitle}{Free energy calculations: {Theory} and applications
  in chemistry and biology}}, edited by
  \bibinfo{editor}{\bibfnamefont{C.}~\bibnamefont{Chipot}} \bibnamefont{and}
  \bibinfo{editor}{\bibfnamefont{A.}~\bibnamefont{Pohorille}}
  (\bibinfo{publisher}{Springer}, \bibinfo{year}{2007}),
  vol.~\bibinfo{volume}{86} of \emph{\bibinfo{series}{Springer series in
  chemical physics}}, chap.~\bibinfo{chapter}{9}.

\bibitem[{\citenamefont{Shah et~al.}(2007)\citenamefont{Shah, Asthagiri, Pratt,
  and Paulaitis}}]{shah:144508}
\bibinfo{author}{\bibfnamefont{J.}~\bibnamefont{Shah}},
  \bibinfo{author}{\bibfnamefont{D.}~\bibnamefont{Asthagiri}},
  \bibinfo{author}{\bibfnamefont{L.}~\bibnamefont{Pratt}}, \bibnamefont{and}
  \bibinfo{author}{\bibfnamefont{M.}~\bibnamefont{Paulaitis}},
  \bibinfo{journal}{J. Chem. Phys.} \textbf{\bibinfo{volume}{127}},
  \bibinfo{pages}{144508} (\bibinfo{year}{2007}).

\bibitem[{\citenamefont{Paliwal et~al.}(2006)\citenamefont{Paliwal, Asthagiri,
  Pratt, Ashbaugh, and Paulaitis}}]{PaliwalA:Anamp}
\bibinfo{author}{\bibfnamefont{A.}~\bibnamefont{Paliwal}},
  \bibinfo{author}{\bibfnamefont{D.}~\bibnamefont{Asthagiri}},
  \bibinfo{author}{\bibfnamefont{L.}~\bibnamefont{Pratt}},
  \bibinfo{author}{\bibfnamefont{H.}~\bibnamefont{Ashbaugh}}, \bibnamefont{and}
  \bibinfo{author}{\bibfnamefont{M.}~\bibnamefont{Paulaitis}},
  \bibinfo{journal}{J. Chem. Phys.} \textbf{\bibinfo{volume}{124}},
  \bibinfo{pages}{224502 } (\bibinfo{year}{2006}).

\bibitem[{\citenamefont{Asthagiri et~al.}(2006)\citenamefont{Asthagiri, Pratt,
  and Paulaitis}}]{AsthagiriD.:Rolfsm}
\bibinfo{author}{\bibfnamefont{D.}~\bibnamefont{Asthagiri}},
  \bibinfo{author}{\bibfnamefont{L.}~\bibnamefont{Pratt}}, \bibnamefont{and}
  \bibinfo{author}{\bibfnamefont{M.}~\bibnamefont{Paulaitis}},
  \bibinfo{journal}{J. Chem. Phys.} \textbf{\bibinfo{volume}{125}},
  \bibinfo{pages}{24701 } (\bibinfo{year}{2006}).

\bibitem[{\citenamefont{Beck et~al.}(2006)\citenamefont{Beck, Paulaitis, and
  Pratt}}]{Beck:2006}
\bibinfo{author}{\bibfnamefont{T.~L.} \bibnamefont{Beck}},
  \bibinfo{author}{\bibfnamefont{M.~E.} \bibnamefont{Paulaitis}},
  \bibnamefont{and} \bibinfo{author}{\bibfnamefont{L.~R.} \bibnamefont{Pratt}},
  \emph{\bibinfo{title}{The potential distribution theorem and models of
  molecular solutions}} (\bibinfo{publisher}{Cambridge University Press},
  \bibinfo{address}{Cambridge}, \bibinfo{year}{2006}).

\bibitem[{\citenamefont{Castillo et~al.}(2005)\citenamefont{Castillo, Hadi,
  Balakrishnan, and Sarabia}}]{castillo2005}
\bibinfo{author}{\bibfnamefont{E.}~\bibnamefont{Castillo}},
  \bibinfo{author}{\bibfnamefont{A.~S.} \bibnamefont{Hadi}},
  \bibinfo{author}{\bibfnamefont{N.}~\bibnamefont{Balakrishnan}},
  \bibnamefont{and} \bibinfo{author}{\bibfnamefont{J.~M.}
  \bibnamefont{Sarabia}}, \emph{\bibinfo{title}{Extreme Value and Related
  Models with Applications in Engineering and Science}}
  (\bibinfo{publisher}{John Wiley \& Sons}, \bibinfo{year}{2005}).



\bibitem[{\citenamefont{Matubayasi et~al.}(1994)\citenamefont{Matubayasi, Reed,
  and Levy}}]{MATUBAYASIN:THEOTH}
\bibinfo{author}{\bibfnamefont{N.}~\bibnamefont{Matubayasi}},
  \bibinfo{author}{\bibfnamefont{L.}~\bibnamefont{Reed}}, \bibnamefont{and}
  \bibinfo{author}{\bibfnamefont{R.}~\bibnamefont{Levy}}, \bibinfo{journal}{J.
  Phys. Chem.} \textbf{\bibinfo{volume}{98}}, \bibinfo{pages}{10640 }
  (\bibinfo{year}{1994}).

\bibitem[{\citenamefont{Lazaridis}(2000)}]{Lazaridis:2000p304}
\bibinfo{author}{\bibfnamefont{T.}~\bibnamefont{Lazaridis}},
  \bibinfo{journal}{J. Phys. Chem. B} \textbf{\bibinfo{volume}{104,}},
  \bibinfo{pages}{4964} (\bibinfo{year}{2000}).

\bibitem[{\citenamefont{Ben-Naim and Marcus}(1984)}]{ben-naim:2016}
\bibinfo{author}{\bibfnamefont{A.}~\bibnamefont{Ben-Naim}} \bibnamefont{and}
  \bibinfo{author}{\bibfnamefont{Y.}~\bibnamefont{Marcus}},
  \bibinfo{journal}{J. Chem. Phys.} \textbf{\bibinfo{volume}{81}},
  \bibinfo{pages}{2016} (\bibinfo{year}{1984}).

\bibitem[{\citenamefont{Amadei et~al.}(1996{\natexlab{a}})\citenamefont{Amadei,
  Apol, Nola, and Berendsen}}]{amadei:1560}
\bibinfo{author}{\bibfnamefont{A.}~\bibnamefont{Amadei}},
  \bibinfo{author}{\bibfnamefont{M.~E.~F.} \bibnamefont{Apol}},
  \bibinfo{author}{\bibfnamefont{A.~D.} \bibnamefont{Nola}}, \bibnamefont{and}
  \bibinfo{author}{\bibfnamefont{H.~J.~C.} \bibnamefont{Berendsen}},
  \bibinfo{journal}{J. Chem. Phys.} \textbf{\bibinfo{volume}{104}},
  \bibinfo{pages}{1560} (\bibinfo{year}{1996}{\natexlab{a}}).

\bibitem[{\citenamefont{Amadei et~al.}(1996{\natexlab{b}})\citenamefont{Amadei,
  D, Roccatano, Apol, Berendsen, and DiNola}}]{Amadei:1996p298}
\bibinfo{author}{\bibfnamefont{A.}~\bibnamefont{Amadei}},
  \bibinfo{author}{\bibnamefont{D}}, \bibinfo{author}{\bibnamefont{Roccatano}},
  \bibinfo{author}{\bibfnamefont{M.~E.~P.} \bibnamefont{Apol}},
  \bibinfo{author}{\bibfnamefont{H.~J.~C.} \bibnamefont{Berendsen}},
  \bibnamefont{and} \bibinfo{author}{\bibfnamefont{A.}~\bibnamefont{DiNola}},
  \bibinfo{journal}{J. Chem. Phys.} \textbf{\bibinfo{volume}{105}},
  \bibinfo{pages}{7022} (\bibinfo{year}{1996}{\natexlab{b}}).

\bibitem[{\citenamefont{Hummer et~al.}(1998{\natexlab{b}})\citenamefont{Hummer,
  Garde, Garc\'{i}a, Paulaitis, and Pratt}}]{Hummer98}
\bibinfo{author}{\bibfnamefont{G.}~\bibnamefont{Hummer}},
  \bibinfo{author}{\bibfnamefont{S.}~\bibnamefont{Garde}},
  \bibinfo{author}{\bibfnamefont{A.~E.} \bibnamefont{Garc\'{i}a}},
  \bibinfo{author}{\bibfnamefont{M.~E.} \bibnamefont{Paulaitis}},
  \bibnamefont{and} \bibinfo{author}{\bibfnamefont{L.~R.} \bibnamefont{Pratt}},
  \bibinfo{journal}{Proc. Natl. Acad. Sci. USA} \textbf{\bibinfo{volume}{95}},
  \bibinfo{pages}{1552Ð1555} (\bibinfo{year}{1998}{\natexlab{b}}).


\end{thebibliography}

\end{document}